\documentclass[]{spie}  

\usepackage{amsmath,amsfonts,amssymb}
\usepackage{graphicx}
\usepackage{subcaption}
\usepackage{bm}
\usepackage[numbers]{natbib}
\usepackage{booktabs}
\usepackage{amsmath}
\usepackage[colorlinks=true, allcolors=blue]{hyperref}

\title{An open-source data processing pipeline for Keck / NIRC2-Polarimetry}

\author[a]{Briley L. Lewis}
\author[a]{Maxwell A. Millar-Blanchaer}
\author[a]{Rebecca Zhang}
\author[b]{Jayke Nguyen}
\author[c]{Max Brodheim}
\author[a]{Ashish Uhlmann}
\affil[a]{University of California Santa Barbara, Santa Barbara, CA 91306 USA}
\affil[b]{University of California San Diego, La Jolla, CA 92093 USA}
\affil[c]{W.M. Keck Observatory, Kamuela, HI 96743 USA}

\authorinfo{Further author information: (Send correspondence to Briley L. Lewis)\\E-mail: brileylewis@ucsb.edu}

\begin{document} 
\maketitle

\begin{abstract}
The Keck/NIRC2 infrared imager was recently upgraded with a new suite of polarimetric observing modes. This polarimetry upgrade (referred to as NIRC2-Pol) will open up a wide range of new astronomical studies, including investigations of exoplanets, the Galactic center, active galactic nuclei and the solar system. Astronomical polarimeters require sophisticated data reduction, especially for quantitative polarimetry requiring the use of a Mueller Matrix model for instrumental polarization calibration. This work presents the design of a flexible, user-friendly, open-source data processing pipeline intended for use with NIRC2-Pol, summarizing its key steps and details of its ongoing implementation.  
\end{abstract}


\section{INTRODUCTION}

Polarimetry is useful for many science cases, as polarized light appears from multiple astrophysical phenomena, such as light scattering off of small dust grains or synchrotron radiation (i.e. near strong magnetic fields). It is particularly useful in high-contrast imaging for circumstellar disk science, where polarimetric differential imaging (PDI) enables separation of host star light and scattered light from the disk \citep{kuhn2001imaging,follette2023intro}.

A handful of adaptive optics (AO)-fed visible and near-infrared imaging polarimeters currently exist: VLT SPHERE/IRDIS \citep{de2020polarimetric,van2020polarimetric}, Subaru/SCExAO/CHARIS \citep{gj2021full,lawson2021high}, Gemini/GPI(2.0) \citep{perrin2010imaging,millar2016gpi,chilcote2020gpi}, Subaru/SCExAO VAMPIRES \citep{zhang2023characterizing,lucas2024visible}, Magellan/MagAO-X \citep{de2025concept,lucas2026spie}, and Subaru/IRCS \citep{terada2018thermal,watanabe2018near}. Yet, there is no unified scheme for processing polarimetric data. Each of these instruments has its own methods for polarimetric data reduction and its own polarimetric coordinate definitions, which in many cases differ from the textbook definitions and sign conventions for polarimetric quantities \cite{goldstein2017polarized}. Additionally, each instrument must contend with observatory-specific coordinate choices, optical paths/designs, data formatting, \texttt{FITS} header keywords, and more.
In this work, we do \textit{not} attempt to create one data processing pipeline to rule them all---instead, we present the framework for a user-friendly, open-source data processing pipeline for Keck/NIRC2-Polarimetry (a.k.a. NIRC2-Pol), a newcomer to the roster of AO-fed infrared polarimeters for astronomy. 

NIRC2 is a near-infrared imager ($\sim$1 to 5 $\mu$m, \textit{Y} through \textit{M} bands) at the W.M. Keck Observatory's 10 meter Keck II telescope, positioned behind the telescope’s facility adaptive optics system \citep{van2004performance,wizinowich2006wm,lilley2024keck}. As a truly multi-purpose imager, it contains multiple filter wheels with a selection of pupil masks, wavelength filters, grisms, spectroscopic slits, coronagraphic spots, and more \citep{castella2016commissioning, xuan2018characterizing}. As of October 2025, NIRC2 has been upgraded to enable dual-channel polarimetry observations, via the addition of three new optics: a Wollaston prism (polarizing beamsplitter), a 5x10'' field mask to prevent overlap between the ordinary and extraordinary beams, and a half-wave plate (HWP) to modulate the signal for calibration \citep{lewis_2026_20737935,lewis2026nirc}. Observations are taken in sequences of four ``critical angles'' to retrieve Stokes $I$, $Q$, and $U$. This upgrade began in 2019 with the installation of the Wollaston prism in NIRC2, and was completed with the HWP installed in August 2025; commissioning and science verification are nearing completion, and the mode first became available to the community in the recent 2026B call for proposals. NIRC2-Pol is a unique instrument: Keck II is the largest telescope on which AO-fed IR polarimetry is available; the instrument spans a large wavelength range including unique access at \textit{L'}; and the mode can work in conjunction with a number of other features already available in the system, including both natural and laser guide star (NGS/LGS) AO. It is notably the only polarimeter equipped with a vortex coronagraph, presenting a unique opportunity to combine these two technologies \citep{feinberg2024habitable} as well as an interesting data processing challenge to be tackled in forthcoming work from the NIRC2-Pol team \citep{mbprepMWC}.

Historically, open-source data processing options for NIRC2 have been limited or tailored to specific uses: \texttt{AIR.jl} can handle simple preprocessing (i.e. dark subtraction, flat fielding) and is written in Julia \citep{Nguyen}; the Keck AO Imaging (KAI) pipeline is based in IRAF and not intended for high-contrast/coronagraphic data \citep{lu_2021_6677744}; and \texttt{nirc2\_reduce} was designed specifically for twilight solar system observations. Although these are highly useful tools, there is no ``one-stop-shop'' for dealing with NIRC2 data---especially not in the new polarimetric mode. Similarly, there are no existing general polarimetric data reduction packages that can easily port to handle NIRC2-Pol data (although some existing packages, particularly \texttt{pyMuellerMat} \citep{max_millar_blanchaer_2026_20752256} and \texttt{pyPolCal} \citep{thomas_mcintosh_2026_20752634}, can handle parts of the functionality needed for a polarimetric pipeline).

As polarimetry is already quite a challenging observing mode, we aim to provide open-source, easy-to-use data processing tools that can make the new NIRC2-Pol mode more accessible to an observer. This pipeline is not currently planned to be connected to the Keck Observatory Archive for automatic data reduction; instead reduction is the responsibility of the observer. From pre-processing to Stokes cube calculations, the \texttt{NIRC2Pol-DPP}\footnote{https://github.com/UCSB-Exoplanet-Polarimetry-Lab/NIRC2Pol-DPP} (Data Processing Pipeline) is intended to both quickly and easily provide a rough reduction of NIRC2-Pol data with minimal need for user input, while also having options for a more experienced user to tinker with the reduction parameters and choices to improve the resulting output data. 

Additionally, a guiding principle of this software's design was its potential for expansion to other instruments, to reduce duplication of effort across similar polarimetric instruments. As a result, this pipeline is designed to be compatible with and developed in conjunction with forthcoming generalized polarimetric imaging packages, which will host core functionality common to these many imaging polarimeters. Further details on the design decisions and workflow of the pipeline are discussed in Section \ref{sec:design}, while an example of the data processing is shown in Section \ref{sec:example}.

\section{ARCHITECTURE AND DESIGN GOALS}\label{sec:design}

\begin{figure*}[h!]
    \centering
    \includegraphics[width=0.75\linewidth]{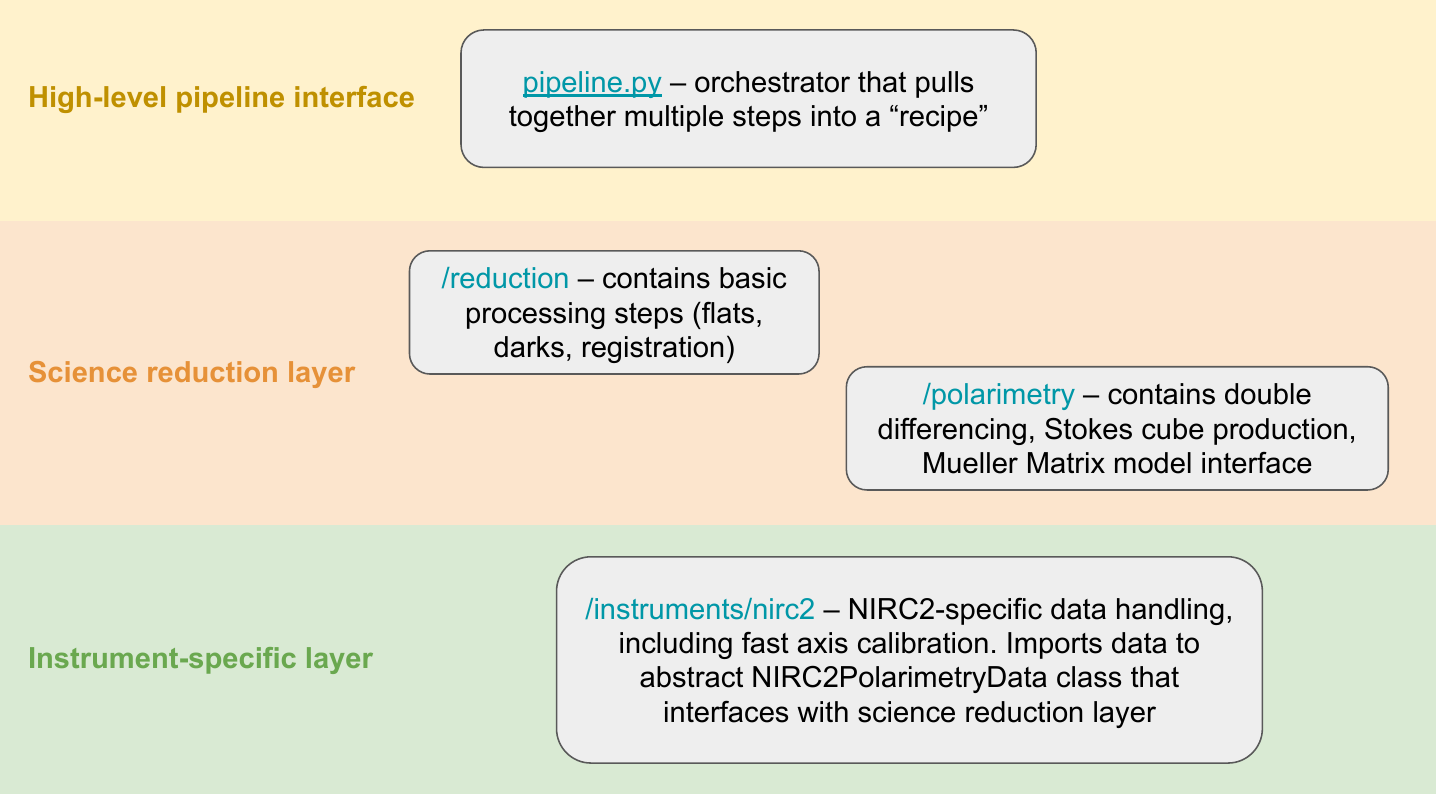}
    \caption{The NIRC2-Pol DPP abstracts the science reduction and polarimetric calculations from instrument-specific I/O. This diagram shows the layers of abstraction: the instrument-specific base layer, the science reduction layer, and the high-level interface to the pipeline. A user may choose to engage with the high-level pipeline orchestrator, or to create a custom reduction using modules from the science reduction layer.}
    \label{fig:abstraction}
\end{figure*}

The design choices herein were inspired largely by \texttt{pyKLIP}'s modularity and abstraction \citep{wang2015pyklip}, allowing for flexibility in the level of user engagement. Similar to \texttt{pyKLIP}, we also introduce layers of abstraction to separate the instrument-specific details from the core scientific functionality, as described in Figure \ref{fig:abstraction}. The instrument-specific layer handles all file I/O (such as parsing \texttt{FITS} headers) and imports relevant calibration values, containing this information in an abstract \texttt{PolarimetryData} class. This layer then interfaces with the science reduction steps, which are agnostic to the instrument input; in future work, this layer may be replaced by a separate generalized package for polarimetric imaging reduction and calibration, which is currently in early stages of development. These layers are combined by a higher-level pipeline interface layer, which orchestrates the workflows into a user-friendly, nearly push-button quick reduction of the data.

Simultaneously, each of the steps used in that pipeline orchestrator are highly modular, with various options that a user can toggle to adjust the reduction for their specific dataset. For example, an experienced user may wish to perform image registration one of several ways; those options are exposed to the user, who can therefore tweak that step of the recipe. Similarly, if a user would like to iteratively improve on their image centering by inspecting radial Stokes cubes, they could do so by creating a script using existing functions. 

\begin{figure*}
    \centering
    \includegraphics[width=\linewidth]{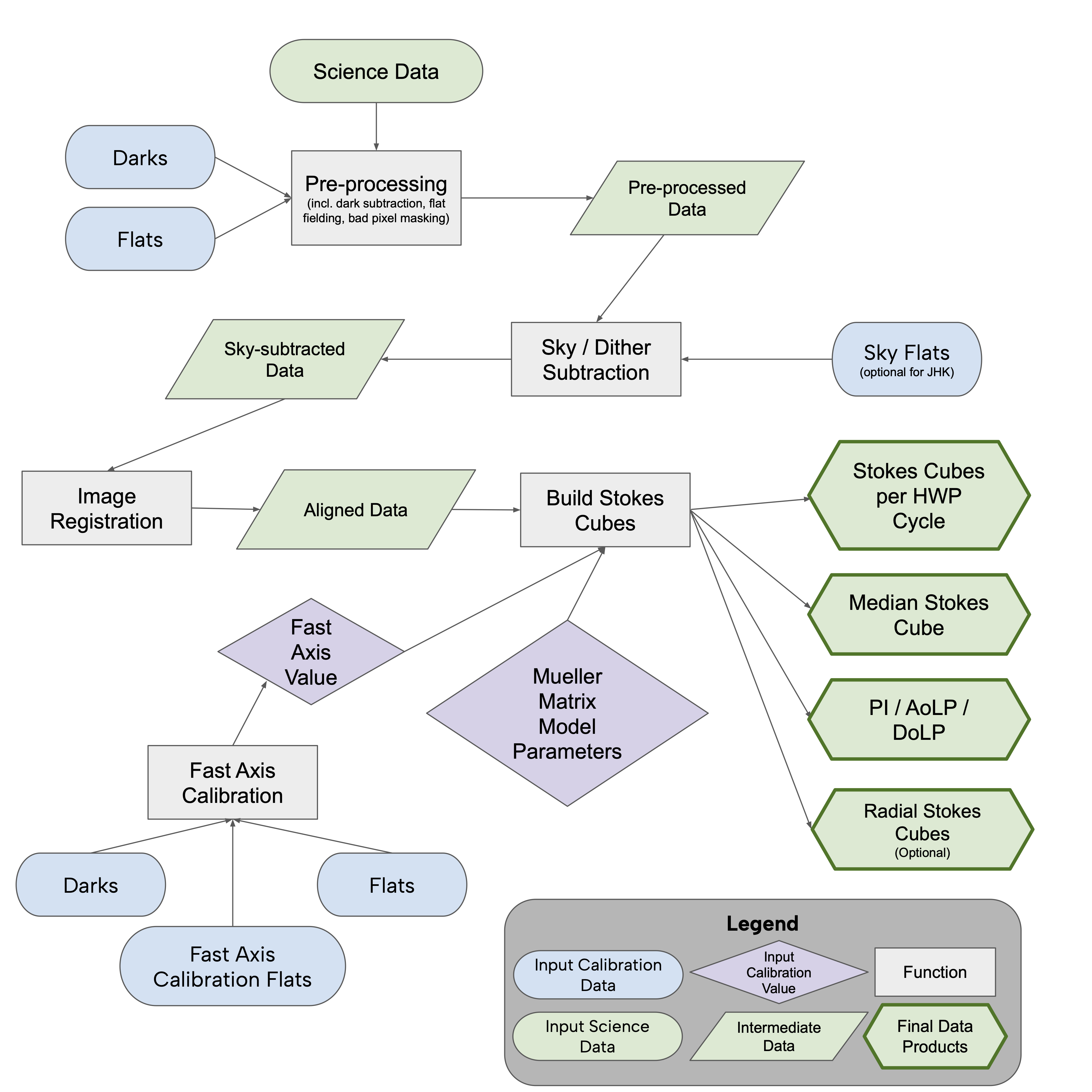}
    \caption{Block diagram of inputs and outputs for the NIRC2-Pol DPP. Science data and related products are indicated in green, with ovals for input data, parallelograms for intermediate data products, and hexagons for the final data products. Calibration imaging data (e.g. darks, flats) are indicated in blue ovals, while other calibration values are indicated by purple diamonds. Functions are denoted by grey rectangles. The final products of this DPP are Stokes cubes per HWP cycle, a median-collapsed Stokes cube, derived products (PI: polarized intensity, DoLP: degree of linear polarization, AoLP: angle of linear polarization), and optionally radial Stokes cubes. Fast axis calibration files are optional, as a user may choose to rely on other recently computed calibration values, and the Mueller matrix model is described in \cite{zhang2026spie}.}
    \label{fig:block}
\end{figure*}

A block diagram of the package's high-level functionality, including inputs and outputs, is shown in Figure \ref{fig:block}. A variety of calibration data are necessary for NIRC2-Pol, including darks, flats, fast axis calibration data, and Mueller matrix model parameters. Details of the Mueller matrix model calibration's incorporation into the data processing are not yet available, as the system model is still currently under development \citep{zhang2026spie}. Details of the pipeline's workflow (with a placeholder for the Mueller matrix model calibration) are described in the following section, along with illustrative examples of the data at each stage.

\section{WORKFLOW AND EXAMPLE DATA}\label{sec:example}

\begin{figure*}
    \includegraphics[width=\linewidth]{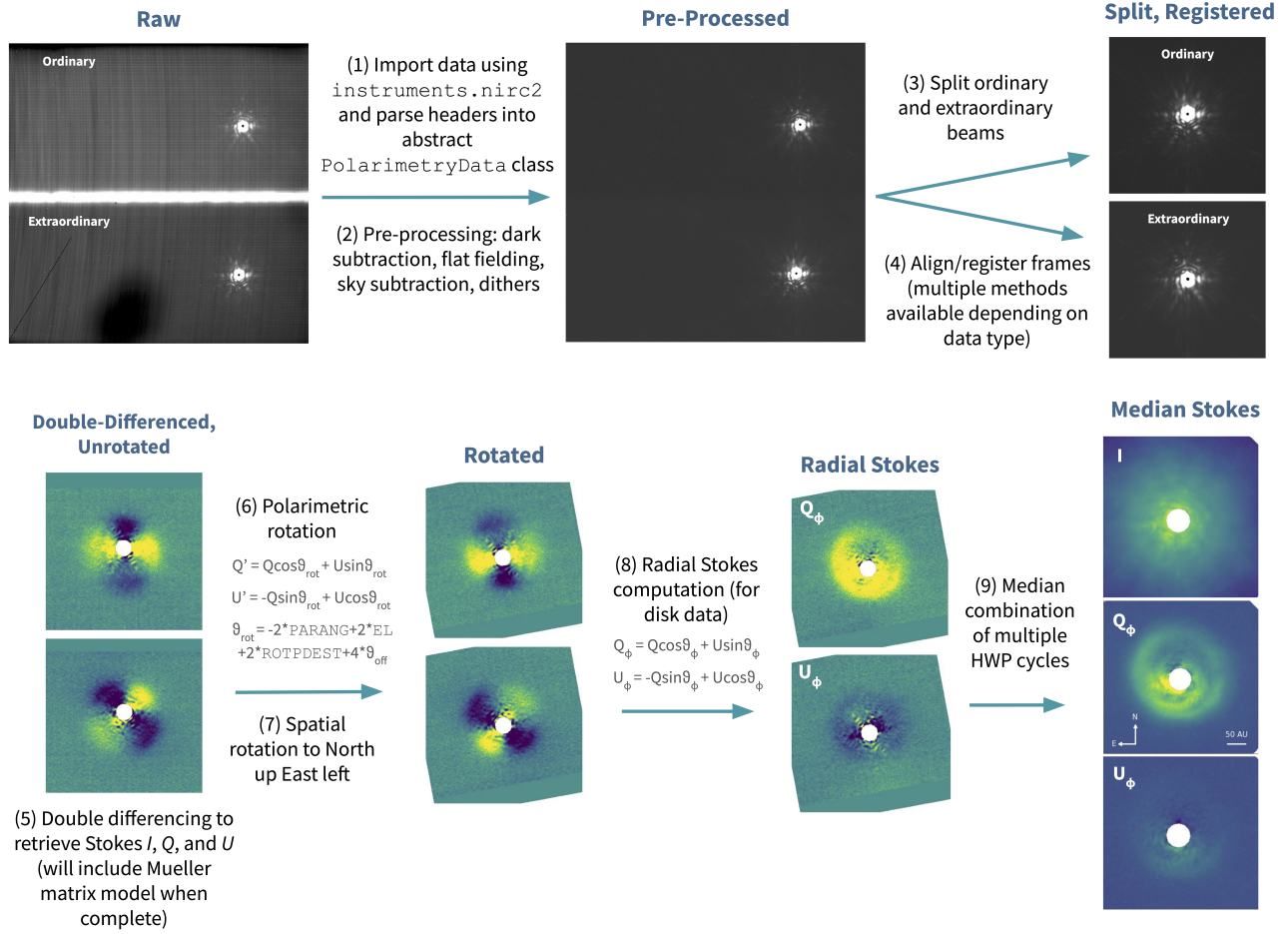}
    \caption{(Top) A single frame at a single HWP angle from raw to pre-processed (dark subtraction/flat fielding), then split into the ordinary and extraordinary beam and registered, making it ready for double differencing. (Bottom) Combinations of frames at multiple HWP angles, illustrating double-differencing, rotation, and conversion to ``radial'' Stokes, a useful quantity for disks. In the resulting median Stokes image, you can clearly see the spirals of AB Aur's protoplanetary disk in $Q_\phi$.}\label{fig:abaur}
\end{figure*}

In this section, we walk through step-by-step a typical reduction, illustrated by the protoplanetary disk AB Aurigae in Figure \ref{fig:abaur} \citep{boccaletti2020possible,Currie2022NatAs,Bowler2025AJ}, as observed during commissioning for NIRC2-Pol in December 2025 \citep{lewis2026nirc}.

\subsection{Data Import and Pre-Processing}

First, data must be imported and relevant header keywords pulled. For polarimetry, it is critical to extract the HWP angle (in this case for Keck/NIRC2-Pol, the value \texttt{PCUPR}). Other polarimetrically relevant quantities exist in the Keck \texttt{FITS} headers, such as \texttt{PCUNAME} which indicates the named position of the PCU2 which holds the HWP \citep{claveau2026inprep}. Existing pre-processing software such as \texttt{AIR.jl}\footnote{https://github.com/jsnguyen/AIR.jl} \citep{Nguyen} automatically matches darks and flats to the necessary data based on exposure times and detector settings; we plan to include similar functionality. Likewise, the most recent fast axis angle calibration will be pulled from a .csv file record of previously completed calibrations based on the date of the observations, but can also be overridden by the user to use the calibration of their choice.

Although some instruments, such as Gemini/GPI, combine HWP angles before Matrix inversion \citep{perrin2014polarimetry,perrin2015polarimetry,millar2016gpi,wiktorowicz2014gemini}, many of the science cases of interest from NIRC2-Pol commissioning require time resolution on the scales of one HWP cycle (0$^\circ$, 45$^\circ$, 22.5$^\circ$, 67.5$^\circ$), necessitating a so-called ``HWP matching'' step; this is similar to that of the CHARIS-DPP \citep{currie2020sky} where HWP angles are grouped together into cycles of the necessary critical angles and the mapping of frame number to HWP cycle number is recorded. The pipeline also provides the flexibility to combine HWP cycles as desired, and in the future, may also include a GPI-style reduction option.

\subsection{Sky and Dither Subtraction}

Sky subtraction is recommended using typical annulus methods from aperture photometry (i.e. \texttt{photutils} \citep{photutils}) for \textit{J}, \textit{H}, \textit{K} bands, and that is typically sufficient even without dedicated sky frames for \textit{J}, \textit{H}, and often also \textit{Kp}. For \textit{L'}, however, the thermal background changes rapidly and is highly spatially variable \citep{nguyen2025ground}. As a result, for most \textit{L'} observations, it is necessary to dither and perform an additional subtraction here between the two positions. Both possibilities are included in the pipeline.

\subsection{Image Registration}

After splitting the image into the ordinary and extraordinary beam to make a data cube with two slices, we perform image registration. There are a number of ways to center an image: e.g. cross-correlation, fitting a gaussian, searching for the maximum pixel at the core, searching for a \textit{minimum} pixel near the core if it is saturated, and more. The software is designed such that a user can choose one of these options or ``drop-and-replace'' their favorite registration algorithm in a fairly straightforward manner.

\subsection{Fast Axis Calibration}

Before proceeding to Stokes cube generation, it is necessary to determine the offset angle of the HWP fast axis from the measured zero position of the rotation stage. Although this value should be repeatable with high precision, we are currently re-testing periodically to ensure this is true during commissioning. The pipeline includes a routine to analyze a so-called ``fast axis calibration sequence'' where the HWP is rotated from 0 to 180$^\circ$ in increments of 10 while observing a calibration source. This process and its justification are further described in \citep{lewis2026nirc}. The result of this code is a single floating point value ($\theta_{\rm off}$) that will be used in the polarimetric rotation during Stokes cube generation. Each time a fast axis calibration is performed by the instrument team, new values will be added to a log file in the pipeline, from which the most recent fast axis calibration can be automatically pulled for an observer's data reduction; there is also the option for the user to run this procedure again and update the value.

\subsection{Stokes Cube Generation}

In this section, we assume an idealized system; further updates to this simplistic handling of instrumental polarization effects will be made in the near future with the Mueller matrix modeling work in \citep{zhang2026spie}. To retrieve Stokes \textit{Q} and \textit{U} in the instrument frame, we take double differences, where $I_{\rm top}(\theta)$ and $I_{\rm bottom}(\theta)$ are the two orthogonal polarization states on each half of the detector (i.e. the ordinary and extraordinary beams) for a given HWP angle, and $Q^+$/$U^+$ and $Q^-$/$Q^-$ are the single differences defined as follows:
\begin{equation}
\begin{split}
    Q &= \frac{1}{2}(Q^+ - Q^-) \\
      &= \frac{1}{2}(I_{\rm top}(0^\circ) - I_{\rm bottom}(0^\circ))  - (I_{\rm top}(45^\circ) - I_{\rm bottom}(45^\circ))
\end{split}
\end{equation}
\begin{equation}
\begin{split}
    U &= \frac{1}{2}(U^+ - U^-) \\
      &= \frac{1}{2}(I_{\rm top}(22.5^\circ) - I_{\rm bottom}(22.5^\circ))  - (I_{\rm top}(67.5^\circ) - I_{\rm bottom}(67.5^\circ))
\end{split}
\end{equation}
Double-differencing in this manner removes instrumental polarization downstream of the HWP. If we assume the many Mueller matrix components in the beam are ideal, we can simplify this to basic equations for rotation. In terms of the relevant NIRC2 header keywords, the overall polarimetric rotation is then: 
\begin{equation}
    \theta_{\rm rot} = -2\times\texttt{PARANG} + 2\times\texttt{EL} + 2\times\texttt{ROTPDEST} + 4\times\theta_{\rm off}
\end{equation}
using the following equations to relate the measured $Q$ and $U$ to those in the sky frame $Q'$ and $U'$:
\begin{equation}
    Q' = Q{\rm cos}\theta_{\rm rot} + U{\rm sin}\theta_{\rm rot}
\end{equation}
\begin{equation}
    U' = -Q{\rm sin}\theta_{\rm rot} + U{\rm cos}\theta_{\rm rot}
\end{equation}
where \texttt{PARANG} is the parallactic angle, \texttt{EL} is the elevation, \texttt{ROTPDEST} is the image rotator position, and $\theta_{\rm off}$ is the fast axis offset angle derived earlier. We note that \texttt{ROTPDEST} is 2 times the physical angle of the rotator to the bench (denoted as \texttt{OBRT}), so it holds a factor of two, \textit{not} four like the fast axis offset angle. After the polarimetric rotation, we must spatially rotate the images---specifically, we are de-rotating the sky-frame Stokes images $Q'$ and $U'$ frames to North up - East left using \texttt{pyklip.rotate} \citep{wang2015pyklip,Wang2015ascl}.

Note that there are multiple optional outputs from the Stokes cube generation step of the pipeline. One can have individual Stokes cubes per HWP cycle, a median-combined Stokes cube, derived quantities such as polarized intensity (PI), angle of linear polarization (AoLP), and degree of linear polarization (DoLP), and/or radial Stokes which is defined as follows:
$Q_\phi$ and $U_\phi$ are the radial Stokes parameters, where $\phi$ is the azimuthal angle around the disk (-$\pi$ to $\pi$ measured counterclockwise from -x); in this frame, $Q_\phi$ contains disk signal and $U_\phi$ should contain noise:
\begin{equation}
    Q_\phi = Q{\rm cos}(2\theta) + U{\rm sin}(2\theta)
\end{equation}
\begin{equation}
    U_\phi = -Q{\rm sin}(2\theta) + U{\rm cos}(2\theta)
\end{equation}

\section{CONCLUSIONS}
\label{sec:conclusion}

The NIRC2Pol-DPP is actively under construction, expected to reach its first release in Fall 2026. A variety of different data from the NIRC2-Pol Science Verification program taken in Semester 2026A (point sources, extended objects, crowded fields near the galactic center, non-sidereal objects) are being used to stress-test the pipeline to ensure its flexibility. Simultaneously, the full Mueller matrix model of the system is under development \citep{zhang2026spie} and will be incorporated into the data processing pipeline as it becomes available, enabling quantitative polarimetry with this instrument. This pipeline's design focuses on modularity and simplicity, keeping individual steps isolated to encourage the most flexibility for future use while also prioritizing ease-of-use for a general observer. Future work beyond the pipeline's first implementation will rely on and contribute to generalized packages for polarimetric imaging data processing and calibration, with the vision of creating a set of tools useful for a wide range of polarimetric instruments. 

\acknowledgments 
 
B.L.L. acknowledges support from the National Science Foundation Astronomy \& Astrophysics Postdoctoral Fellowship under Award No. 2401654. Any opinions, findings, and conclusions or recommendations expressed in this material are those of the author(s)and do not necessarily reflect the views of the National Science Foundation. 

Some of the data presented herein were obtained at Keck Observatory, which is a private 501(c)3 non-profit organization operated as a scientific partnership among the California Institute of Technology, the University of California, and the National Aeronautics and Space Administration. The Observatory was made possible by the generous financial support of the W. M. Keck Foundation. This research has made use of the Keck Observatory Archive (KOA), which is operated by the W. M. Keck Observatory and the NASA Exoplanet Science Institute (NExScI), under contract with the National Aeronautics and Space Administration. The authors wish to recognize and acknowledge the very significant cultural role and reverence that the summit of Maunakea has always had within the Native Hawaiian community. We are most fortunate to have the opportunity to conduct observations from this mountain. 

\bibliography{report} 
\bibliographystyle{spiebib} 

\end{document}